# Clustering Wi-Fi Fingerprints for Indoor-Outdoor Detection

Mr. Guy Shtar, Prof. Bracha Shapira, Prof. Lior Rokach

shtar@post.bgu.ac.il, bshapira@bgu.ac.il, liorrk@post.bgu.ac.il

Corresponding author: Mr. Guy Shtar, shtar@post.bgu.ac.il, Telekom Innovation Laboratories at Ben-Gurion University of the Negev, P.O.B. 653, Beer-Sheva, 84105, Israel

**Abstract**: This paper presents a method for continuous indoor-outdoor environment detection on mobile devices based solely on Wi-Fi fingerprints. Detection of indoor-outdoor switching is an important part of identifying a user's context, and it provides important information for upper layer context aware mobile applications such as recommender systems, navigation tools, etc. Moreover, future indoor positioning systems are likely to use Wi-Fi fingerprints, and therefore Wi-Fi receivers will be on most of the time. In contrast to existing research, we believe that these fingerprints should be leveraged, and they serve as the basis of the proposed method. Using various machine learning algorithms, we train a supervised classifier based on features extracted from the raw fingerprints, clusters, and cluster transition graph. The contribution of each of the features to the method is assessed. Our method assumes no prior knowledge of the environment, and a training set consisting of the data collected for just a few hours on a single device is sufficient in order to provide indoor-outdoor classification, even in an unknown location or when using new devices. We evaluate our method in an experiment involving 12 participants during their daily routine, with a total of 828 hours' worth of data collected by the participants. We report a predictive performance of the AUC (area under the curve) of 0.94 using the gradient boosting machine ensemble learning method. We show that our method can be used for other context detection tasks such as learning and recognizing a given building or room.

## 1.     Introduction

Context aware systems are able to sense their environment and adapt accordingly [1]. Such systems can be used to deliver relevant information and services to a user. Mobile devices contain numerous sensors that enable accurate identification of a user's context [2], and location is a critical dimension of context. Today's applications allow us to find a nearby restaurant, obtain directions to a location, and order a taxi. The ability to position devices and people in all types of settings seamlessly, accurately, and globally will lead to even more innovative applications and business opportunities.

While global navigation satellite systems (GNSSs) have advanced considerably, indoor positioning systems (IPSs) remain a major technological challenge. Moreover, IPSs and GNSSs rely on different sensors, and it remains difficult for today's devices to determine which sensors to use at any given time. Knowing whether the user is indoors or not can prevent systems from activating unneeded sensors, thereby conserving device resources such as the battery and CPU (e.g., prevent the activation of GPS indoors), and enable algorithms to make wiser decisions. Recommender systems, activity recognition, logical localization, and automatic image annotation will benefit from improved context information.

Previous work in the field points to the need for indoor-outdoor detection. Nevertheless, none of the prior research has focused on Wi-Fi fingerprints for indoor-outdoor detection. The number of Wi-Fi access points (APs) is growing, and the characteristics associated with Wi-Fi APs (e.g., higher density around buildings) represent resources that could be harnessed specifically for indoor-outdoor detection. Despite the fact that future mobile devices will likely scan continuously for Wi-Fi fingerprints to serve IPSs, previous work has not focused on the Wi-Fi receiver and has instead focused on approaches that involve the use of unnecessary sensors and thus result in increased battery usage.

Our research centers on Wi-Fi fingerprints and how they should be clustered, and more importantly, how they might best be leveraged, to classify the environment for indoor-outdoor detection. Although we don't tackle the issue of power consumption directly, the presented method can be optimized to reduce power consumption (e.g., by adjusting the scan rate based on the amount of motion using the accelerometer).

While some past research has dealt with clustering Wi-Fi fingerprints on a small scale (e.g., in a specific building) or has dealt with indoor-outdoor detection without using Wi-Fi fingerprints, in our research we propose a method for indoor-outdoor detection based on Wi-Fi fingerprinting; we also provide our rationale for doing so and describe the experiment conducted in order to evaluate the method.

In order to evaluate the proposed method, an Android application was developed to collect the Wi-Fi fingerprints and ground truth values. This is critical as using a dataset that fails to describe the behavior of the users or using inappropriate evaluation criteria for evaluating an indoor-outdoor detection method can be misleading as indoor-outdoor detection is somewhat imbalanced.

While some existing methods are aimed at detecting whether the user is located either indoors or outdoors, these methods do not generalize well to unknown environments or devices; furthermore, they fail to obtain a sufficient level of accuracy and do not rely on sensors that are likely to be activated (such as Wi-Fi sensors). The method presented in this research is expected to lead to continuous indoor-outdoor environment detection on mobile devices. The contributions of this research are fourfold:

1. Designing a new method for training an indoor-outdoor classifier with the following three properties: (1) global coverage - while the classifier is trained on a specific location, it can still be generalized to unknown locations (no need to create a radio map), (2) battery efficient - it utilizes just the Wi-Fi receiver, (3) device agnostic - it supports any mobile device, regardless of the model.
2. Introducing the notion of a transition graph for Wi-Fi fingerprints, which is useful for extracting informative features for the indoor-outdoor classification task.
3. Conducting a field study for collecting real data (the dataset was made publicly available).
4. Demonstrating that rank correlation coefficients are useful as distance measures for Wi-Fi fingerprints and presenting a new index structure for efficient distance calculations.

This paper is organized as follows. Section 2 contains a review of previous work in the field. Section 3 describes the proposed methods, including Wi-Fi fingerprint clustering, transition graph creation, and indoor-outdoor detection. The measurement and evaluation methods are described in section 4. Section 5 is dedicated to presenting the results. We sum up our research and discuss the results and limitations in section 6, and finally, the conclusions and future work are presented in section 7.

## 2. Related Work

This research encompasses the fields of data mining, indoor positioning and mapping using Wi-Fi fingerprints; this section begins with a review of existing indoor-outdoor detection systems, and continues with a review of indoor positioning and mapping, and logical localization. The section ends with a review of existing distance measures.

### 2.1. Indoor-Outdoor Detection

Several recent works have attempted to deal with indoor-outdoor detection. BlueDetect [3] is based on the emerging iBeacon technology and excels in its low power consumption but relies on the installation of specific hardware on-site. Wang et al. [4] uses the Global System for Mobile (GSM) communication cellular base station's signal strength to attack the problem and achieves high accuracy, however the method was only tested using a single device. Ravindranath et al. uses the GPS receiver to deal with the problem, and their main contribution is demonstrating how the output of such a system can be used to improve wireless network performance [5]. Cho et al. rely solely on the GPS receiver for indoor-outdoor detection [6]. As described in the research referenced below, indoor-outdoor detection using GPS suffers from high power usage and can be biased in outdoor, highly-dense urban areas, where a clear line of sight with the satellites is unavailable. Various methods use multiple sensors such as the accelerometer, proximity and light sensors, time, GSM receiver, and magnetometer [7, 8]. Radu et al. provide a review of the available techniques and critique them, generally arguing that GPS based methods are both inaccurate and high power consumers [9]. In particular, they were critical of the IODetector for its hard-coded thresholds, stating that is unable to function in new places. Their main contribution is the use of semi-supervised training: clustering, self-training, and co-training for indoor-outdoor detection. WifiBoost [10] relies solely on Wi-Fi APs, however it is meant to work after a preliminary survey, making it a local solution that cannot operate globally. Edelev et al. focus on indoor-outdoor detection using the GPS, a weather web service, and sound data [11]. Indoor-outdoor detection of predefined regions is dealt with by He et al. [12], and the authors also discuss the issue of calibrating heterogeneous devices. Anagnostopoulos and Deriaz [13] use a simple algorithm based on comparing the error estimation of indoor and outdoor positioning systems; the algorithm can operate with any technology that supports error estimation.

The main drawback of many of the works presented in this subsection is that they reported the accuracy but did not specify the AUC, or even the recall or precision. Estimates regarding the amount of time that humans spend indoors range from 70 to 90%. Our experiment supports that claim, showing that the average user spends 83% of the time indoors. Indoor-outdoor detection is therefore somewhat imbalanced, and thus it should be evaluated appropriately. Evaluation based on accuracy alone is insufficient, as the target attribute (classification) is skewed. Classifying all of the instances as "indoors" would result in an accuracy of 0.83 (in our case). Precision and recall are sometimes used, yet precision is sensitive to data distribution. The AUC is a suitable measure for binary classifier performance. It enables a comparison of the average performance of different classification models and avoids the strong effect of bias in the target attributes [14, 15, 16].

## 2.2. Indoor Positioning and Logical Localization

The popular and widely available GPS has generated interest in a comprehensive indoor positioning service. RADAR [17] paved the way for Wi-Fi based indoor positioning when it showed that a receiver's location could be calculated using pervasive radio frequency beacon signals. Improvements have been made to the technique [18, 19]. UnLoc [20] uses the inertial sensors in smartphones and crowdsourcing in order to avoid a site survey. IONavi [21] suggests a joint navigation solution which operates both indoors and outdoors. IONavi uses crowdsourcing to recommend users the best path using previously collected walking traces. Locus [22] is an indoor localization tracking and navigation system for multi-story buildings; the authors discuss relevant RSS signal transformation and signal differentiation.

Besides delivering accurate geometry (a coordinate) for a location, some works are aimed at classifying logical locations. WILL [23] uses fingerprint clustering to map the indoor environment. Automatically checking in the user at a point of interest using Wi-Fi fingerprints was researched by Bisio et al. [24]. Dousse, Eberle, and Mertens cluster Wi-Fi fingerprints in order to learn and detect places [25]. AWESOM [26] deals with clustering Wi-Fi fingerprints using self-organizing maps.

## 2.3. Wi-Fi Scan & Distance Measures

Wireless local area networks (WLANs) based on IEEE 802.11 specifications periodically transmit a "beacon frame" which announces the presence of the AP. Today's mobile operating systems, such as iOS and Android, allow mobile devices to programmatically "scan" the available Wi-Fi APs located in the area of the device; such scan results can be used for positioning purposes. The results contain the basic service set identification (BSSID) of the AP, although actually the media access control (MAC) address is used, along with its service set identifier (SSID) which is the network name, the received signal strength indicator (RSSI), and other information. The result of such a scan is referred to as a Wi-Fi fingerprint. The RSSI is an important aspect of a fingerprint. The RSSI is a measure of the power in the received radio signal, and it is provided for each AP that was detected during the scan. The RSSI is affected by numerous environmental factors and is considered to be a very noisy signal; the indoor environment (including walls, furniture, etc.), in particular, can cause noise. Nevertheless, a great deal of important research and algorithms are based on the RSSI, and mobile OSs rely on it to position a device in space. The noise factors mentioned above are considered some of the shortcomings associated with indoor positioning, although they can also contribute to indoor-outdoor detection using Wi-Fi as they only characterize indoor environments.

Our method assigns the fingerprints to clusters and creates a cluster transition graph; the clusters and the graph are used later to classify each of the fingerprints. Previous works used a variety of distance measures for Wi-Fi fingerprints during the clustering process. A summary of the distance measures used in previous research is provided by Dousse et al. [25]. WILL used the Euclidean distance [23], and SensLoc used the Tanimoto coefficient which ignores the RSSI and considers the list of APs as a set [27]. LiFS [28] used the Manhattan distance. Rank correlation aims to measure the relationship of different ordinal variables. A distance measure based on ranking was presented by Bisio et al. [24] who formulated a new distance measure and used it to automatically check-in users at point of interests. We suggest using a well-formulated rank based measure over a newly formulated one. Such a rank based fingerprint algorithm for indoor positioning was proposed by Machaj, Brida and Piché [29]; the authors investigated the impact of various rank based distance measures on the positioning accuracy and present a comprehensive report.

## 3. Suggested Framework and Methodology

Our research is motivated by the pioneering works presented in the previous section; we use some of the techniques presented in prior research and incorporate our own innovative ideas to form a new method. This section presents our proposed method, and it includes a formal definition of Wi-Fi fingerprints which rely on the RSSI and a distance measure based on the rank correlation coefficient. The clustering process used is also presented, followed by a description of the creation of the cluster transition graph which is used to extract meaningful features. Finally, we demonstrate how a classifier can be trained using the extracted features.

The purpose of our method is to accurately perform binary classification for indoor-outdoor detection, classifying each fingerprint as one of two labels: indoors or outdoors.

### 3.1. Wi-Fi Fingerprints

Some preliminary definitions are required before describing our method.

A fingerprint matrix $F^D$ is computed from the data gathered by each device $D$:

$$F^D = \begin{bmatrix} f_{11} & f_{12} & \cdots & f_{1N} \\ f_{21} & f_{tn} & \cdots & f_{2N} \\ \vdots & \vdots & \ddots & \vdots \\ f_{T1} & f_{T2} & \cdots & f_{TN} \end{bmatrix}.$$

The elements of the matrix $f_{tn} \in \mathbb{R}_{>0}$ are the power received from AP $n$ during scan $t$. $f_{tn}$ will be equal to zero if AP $n$ was not received at all during scan $t$. $t \in [1, T]$ refers to a single Wi-Fi scan (each row vector is noted as a *fingerprint*); the fingerprints are sorted in the order in which they were collected - ascending order based on the scan start time. $T$ is the total number of fingerprints collected. $n \in [1, N]$ refers to a single AP, and $N$ is the total unique number of APs that were detected during the collection period. As mentioned in the Related Work section, one of the main aspects of a fingerprint is the RSSI. Although RSSI units are not defined by the IEEE 802.11 standard, the RSSI is usually provided in a logarithmic scale. The Android OS, for example, uses the dBm (decibel-milliwatts) as its unit. Converting the RSSI from dBm to power which is measured in a linear scale is possible by using:

$$P = 10^{\frac{RSSI}{10}}. \tag{1}$$

The transformation helps reduce the impact of very weak signals.

Fingerprint $f_t$ is noted as an "empty fingerprint" if it was created from an empty Wi-Fi scan:

$$\forall n \in [1, N]\ f_{tn} = 0. \tag{2}$$

Although two empty fingerprints look exactly the same, it is very likely that they were collected in different locations, unless they were collected sequentially. It is important to note that $F^D$ is sparse, because different APs are available in different locations.

### 3.2. Clustering Wi-Fi Fingerprints

Similar fingerprints are clustered together. Features are extracted from the clusters and later used to classify the fingerprints. The clusters are used as a noise reduction mechanism by averaging the features of similar fingerprints. Additionally, the clusters are later used as the nodes of the transition graph.

#### 3.2.1. Distance Measure

To measure the distance between two fingerprints, we suggest using a well-defined and tested coefficient over a newly formulated one. Such coefficients have been studied and can be tested for significance. Kendall's τ, Spearman's ρ, and Goodman and Kruskal's γ are three common rank correlation coefficients. Our clustering process uses Spearman's ρ [30] as the basis for its distance measure; Spearman's ρ between two random variables, $\vec{X}$ and $\vec{Y}$, is defined as the Pearson correlation between the ranked variables and is computed as follows:

$$\rho(\vec{X}, \vec{Y}) = 1 - \frac{6 \sum d_i^2}{n(n^2-1)}, \qquad (3)$$

where $d_i = rg(x_i) - rg(y_i)$ is the difference between the two ranks (the function $rg(x_i)$ returns the rank of $x_i$), and $n$ is the number of observations. To adapt Spearman's ρ to the sparse nature of $F^D$, we redefine $d_i$ and $n$ as follows:

$$\mathbb{1}(x) = \begin{cases} 1 & if\ x > 0 \\ 0 & otherwise, \end{cases} \qquad (4)$$

$$n = \sum \mathbb{1}(x_i + y_i). \qquad (5)$$

In our case, $n$ is equal to the number of non-zero values in at least one of the variables. We define $\varepsilon > 0$ and smaller than any other non-zero value in $F^D$, and:

$$d_i = \big(rg(x_i + \varepsilon \cdot \mathbb{1}(y_i) \cdot (1 - \mathbb{1}(x_i))) - rg(y_i + \varepsilon \cdot \mathbb{1}(x_i) \cdot (1 - \mathbb{1}(y_i)))\big) \cdot \mathbb{1}(x_i + y_i). \qquad (6)$$

The adaption described above enables the Spearman's ρ to ignore APs that were not received in both of the fingerprints, and ranks APs that were received in just one of the fingerprints last.

Spearman's ρ can be converted from a similarity measure to a distance measure using:

$$\rho_{dist}(\vec{X}, \vec{Y}) = 1 - \rho = \frac{6 \sum d_i^2}{n(n^2-1)}. \qquad (7)$$

The distance measure between two fingerprints $f_i$ and $f_j$ is defined as follows:

$$d_{1-\rho}(f_i, f_j) = \begin{cases} 0 & if\ \forall n \in [1, N]\ (f_{in} = 0 \land f_{jn} = 0) \land |i - j| = 1 \\ 2 & if\ \forall n \in [1, N]\ (f_{in} = 0 \land f_{jn} = 0) \land |i - j| \neq 1 \\ 2 & if\ |\{n \in [1, N]\ |\ f_{in} \neq 0\} \cap \{n \in [1, N]\ |\ f_{jn} \neq 0\}| = 0 \\ 0 & if\ \{n \in [1, N]\ |\ f_{in} \neq 0\} = \{n \in [1, N]\ |\ f_{jn} \neq 0\} \land |\{n \in [1, N]\ |\ f_{in} \neq 0\}| = 1 \\ \rho_{dist}(f_i, f_j) & otherwise. \end{cases} \qquad (8)$$

The first two conditions define the distance involving an empty fingerprint. The third condition defines the result for disjoint fingerprints, and the fourth condition defines the result for the undefined cases of $n = 1$ in Spearman's ρ definition. In other cases, the distance measure is used. Note that the triangle inequality doesn't apply to the measures described above. Spearman's ρ value ranges from a perfect correlation of +1 and an opposite correlation of -1, to no correlation with the value of 0. Therefore, $d_{1-\rho} \in [0,2]$.

### 3.2.2. Clustering Process

Density based clustering excels at handling arbitrary cluster shapes and in cases where the number of clusters is unknown. DBSCAN, the highly regarded density based clustering algorithm [31], is used in our method, in which each fingerprint in $F^D$ is associated with one cluster. The clusters will be classified later, and each cluster's classification will be inherited by the fingerprints forming it. Our system adopts DBSCAN's definition of clusters. Intuitively, a cluster is a non-empty group of dense samples with an arbitrary shape.

The offline clustering process of Wi-Fi fingerprints involves two main challenges: (1) Fingerprints can be produced at high velocity—a few seconds for a new fingerprint. (2) The data is highly dimensional in that each AP in the data is a feature (column vector). In order to cluster a large amount of given Wi-Fi fingerprints in a reasonable time span and address the challenges mentioned, we propose two enhancements to the well-known DBSCAN algorithm: (1) Using a lighter variation of DBSCAN (DBSCANLight) which we formulated to permit fewer regionQuery invocations; a regionQuery is a neighborhood query that returns all instances within a given distance (marked as *eps*) from a given instance. For the sake of brevity DBSCANLight is not presented in this article. Naturally, any clustering algorithm can be used, including DBSCAN itself (which provides similar results but is slower). (2) Restructuring the data so that it can handle the Wi-Fi fingerprints' regionQuery operation efficiently based on Spearman's ρ or any other rank correlation coefficient.

#### 3.2.2.1. Data Structure

This subsection presents a new data structure that supports efficient and quick regionQuery execution based on rank correlation coefficients. In order to support an efficient neighborhood query, algorithms such as DBSCAN are intended to be used in conjunction with spatial indices. The fingerprint space is characterized by a large number of instances where the vast majority of instance couples result in a similarity of 0 (maximal distance), as they do not have any AP in common. In order to avoid calculating the distance in such instances, we suggest using mapping, such as a hash table between each AP and the fingerprints that received the AP, as is presented in Fig. 1. For a given fingerprint, we can narrow the neighborhood query to the fingerprints that have at least one AP in common.

So far, we have a list of potential neighbors, but in order to filter the list to the *eps*-distant neighbors, the precise rank correlation coefficient must be calculated. A rank correlation calculation for a fingerprint involves ranking the APs in descending order from the strongest to weakest (note that the ranking for a given fingerprint will probably be used more than once). Therefore, we suggest the use of mapping between each of the APs in a given fingerprint and the AP rank in the fingerprint.

### 3.3. Cluster Transition Graph

In order to consider the order in which the fingerprints were collected—or in other words, the temporal dimension—we suggest using a *cluster transition graph*. The transition graph is used as a map of the wireless networks around the device and serves as an abstraction layer which reduces the difference between different device models. The transition graph $G$ is an undirected and unweighted graph where the nodes are the clusters, and an edge is added between two nodes if the corresponding clusters contain two fingerprints $f_i$, $f_j$, such that $|i - j| = 1$, which indicates that the two fingerprints were collected one after the other. In the resulting graph, outdoor fingerprints tend to line up in long chains in the graph, while the indoor fingerprints tend to remain around a dense area. An interesting observation is that although individuals spend most of their time indoors, the graph contains more outdoor nodes than indoor ones. Fig. 2 shows an example of a transition graph created from one of the datasets collected in this research.

As stated above, indoor and outdoor nodes tend to align themselves differently, and therefore, extracting features from the graph will allow us to classify the nodes of the transition graph, which will further enable us to classify the fingerprints. Moreover, each node represents the fingerprints that formed the corresponding cluster. Hence, the importance of a node, in the sense of the significance of the phenomena it describes, depends on the number of fingerprints that the node represents. As a result of this insight, the number of fingerprints that a node contains (number of fingerprints in the cluster) can be used as a weight for the instances fed to the classifier during the training phase.

Although such a classification algorithm does not rely on explicit input from the user for new locations, some processing is required in order to build a meaningful graph.

### 3.4. Node Features

The transition graph reflects the actual environment abstractly. It carries more information than individual fingerprints or clusters. In order to classify the nodes, the first stage is to extract meaningful features.

The features that were selected to classify a node are those features that are a product of their neighborhood: the number of neighbor nodes, average power in the neighborhood, average number of APs received per scan in the neighborhood, and the average number of fingerprints in the neighborhood. We define the neighborhood of node *x* at distance *d* by:

$$N_x(d) = \{y \mid y \text{ is reachable from } x \text{ at cost } d \text{ or less}\}. \tag{9}$$

The neighborhood is found by returning the result of a breadth-first search from *x*, bounded by distance *d*. The features described above can be calculated for various neighborhood sizes (different values of *d*), and in practice, using various sizes contributes to the final model.

### 3.5. Training a Classifier

This section describes a technique for classifying the fingerprints as indoors or outdoors using the transition graph and supervised machine learning techniques. Each of the nodes is used as an instance, and the features described in subsection 3.4 are used in all of our experiments. The number of fingerprints in each of the nodes is used as its weight. Our method is not limited to a specific classification algorithm. We briefly discuss each of the algorithms evaluated in this research.

Gradient boosting machine (GBM) [32] is an ensemble learning method for regression and classification problems. It is based on the boosting meta-algorithm, which generally consists of iteratively learning weak classifiers and reweighing the misclassified instances; thus, each classifier is focused on instances which were misclassified by previous classifiers. More specifically, GBM learns by consecutively fitting new models to provide a more accurate result. Random forest (RF) [33] is an ensemble learning method that corrects decision trees' habit of overfitting to the training set by "bagging" the instances and creating an unpruned tree from each bag. RF also does this by randomly selecting a subset of the space at each node split. Rotation forest [34] is an ensemble method based on splitting the data into subsets and applying PCA (principal component analysis) to each subset. The naïve Bayes classifier is also used, where Bayes' theorem is applied with the assumption of independence between every pair of features given the class variable. Decision trees with bootstrap aggregation are also used, where multiple random samples (with replacement) are taken from the training dataset, and a classifier is trained from each such random set.

## 4. Measurement and Evaluation

### 4.1. Datasets

In order to evaluate the techniques described in this research, an experiment was conducted involving 12 participants during their daily routine, with a total of 828 hours' worth of data collected by the participants. The participants used their personal mobile phones and were asked to install a specific Android application (Fig. 3) that constantly scanned for Wi-Fi APs and stored the data on the local file system. The users were asked to specify if they were indoors or outdoors whenever they changed location using a checkbox in the application. In the case of an error, the participants had the option to cancel the labels for the last minute, 10 minutes, hour, five hours, or for the last day. The users were asked to participate for a minimum of 24 continuous hours (although one of the users participated for only 22 hours). At the end of the experiment, the data was manually transferred from the devices.

The training set was collected by the research team using the Android application on a Nexus 5 device, and the team's movements and locations were collected in various scenarios. The training data was the only dataset collected with an indication regarding where the data was collected, in order to cross-validate the training set by location (see subsection 4.3 for details). The training data was collected in different types of locations: the user's home, a gym, two additional homes, a small store, research laboratories, and the outdoor environment between these locations, including a few roads of various sizes. The training data collected amounted to about 22 hours' worth of data in total, with approximately 65% of it collected indoors (the rest was collected outdoors). The training set was collected a few months prior to conducting the experiment with the study participants. The test set is composed of the data collected by the experiment's 12 participants. Table 1 presents the training and test data. User 0 represents the data that was collected by the research team which was used to train the model. Each of the participants collected an equal or larger amount of data than the data contained in the training set, with a total of 828 hours' worth of data collected by the participants. Seven different device models were used to allow true cross-device testing. None of the participants used the "cancel last five hours" or "cancel last day" option. The average user spent 83% of the time indoors. The variance among different users was high, as one of the users spent 98% of the time indoors (user 1, a student), and another spent 49% of the time indoors (user 5, a business person). Cluster analysis was performed, and a transition graph was constructed for each of the devices' data separately. The datasets and source code used in this research are available for download at https://goo.gl/voijqC.

### 4.2. Evaluation Measures

To evaluate the indoor-outdoor detection performance, we used the area under the curve (AUC). It is the most appropriate method for evaluating a binary classification model [16], particularly in order to avoid the effect of the bias in the target attributes on the accuracy [15]. The AUC demonstrates the trade-off between the true positive rate (TPR) and the false positive rate (FPR). Accuracy has been used in some indoor-outdoor detection research, but this measure has been shown to be affected by the bias of the target attributes towards one category, as in this case. In our comprehensive evaluation we report the AUC as well as the accuracy and prior probability of the target attribute. Additionally, context switch detection latency in seconds is reported.

Where applicable, we compare the mean AUC and accuracy of the models using a paired t-test and report the two-tailed p-value; we regard the results as significant for p-values below 0.05.

### 4.3. Experimental Setup & Parameter Tuning

All of the experiments were executed on a Lenovo ThinkPad E440 laptop with an Intel i7 processor and 8GB of RAM. The entire framework described in the Suggested Framework and Methodology section was also implemented in Java, and *R* environment was used to train and classify the data.

In general, k-fold cross-validation, where the instances are sampled randomly for each fold, is a good method for tuning and evaluating models. Nevertheless, it is not true to life under these research conditions. For instance, when tuning a classifier using 10-fold cross-validation, the classifier delivers very good results on our training data, but when classifying instances from a new location using the same classifier, the model's performance is inferior. The reason is that once a classifier has been trained on data from a particular place, trying to classify instances nearby in space would result in the identification of the original place. In other words, k-fold cross-validation cannot be used to make accurate conclusions regarding unknown locations. Therefore, to train the model using a single device, location based cross-validation is used instead. The validation is done by leaving instances from one location out when training a model; the model is then tested using the instances that were left out. This process is repeated for each of the locations in the training set, and the final evaluation metric for the training data is the average value – or the average

AUC in this case. For the training set, each node/cluster classification is determined by a majority vote among the labeled fingerprints that formed it. The test set is composed of the data collected by the experiment's 12 participants.

The following parameters were tuned using the method described above. In terms of the clustering process, the MinPts (minimum expected cluster size) is set to 1, so that none of the fingerprints would be classified as noise. The value of *eps* controls the neighborhood size of a given instance for the clustering process, and the value that was selected is *eps* = 0.22. In order to determine the effective distances for defining the neighborhood of a node (the values of *d*), we extract the features across a large distance (our experiments show that a distance of 30 edges is more than enough).

A linear regression model was created, where the dependent variable was the classification itself (indoors/outdoors), and the explanatory variables were the features extracted from each node's neighborhood using the different values of *d*. We determine the significance level for each coefficient (using t-tests) and select the variables that are significant (p-value $\leq 0.05$). Each selected variable represents a different *d* value that will be used as the neighborhood size. The final values for each of the neighborhood sizes are shown in Table 2. A total of 20 features were used—four different features at five different distances (see subsection 3.4). The neighborhood size for the "number of neighbors" feature is unique (distances ranging from 2 to 6); the number of neighbors at a distance of 0 is always equal to 1, and the number of neighbors at a distance of 1 is not statistically significant. To assess the contribution of the clustering and graph creation processes to our system, we train two additional models: (1) a model based on features extracted only from the raw fingerprints, and (2) a model based on features extracted from the clusters. Only applicable features are used; the "number of neighbors" feature isn't used, since it isn't applicable to both of the models. The rest of the features presented in Table 2 are used with a neighborhood size of 0. Each raw fingerprint is considered a single cluster.

## 5. Results

### 5.1. Indoor-Outdoor Detection

Table 3 presents the results for each of the devices (in the 12 rightmost columns), with six columns representing the RF (random forest) model and six representing the GBM (gradient boosting machine); we focus on these models due to their superior AUC score. We report the AUC and accuracy for the models based on the fingerprints, clusters, and transition graph. Additionally, the table shows the devices each user used, demonstrating that there is no difference in the performance of different devices. The table also shows the number of fingerprints collected and the number of clusters created for each device (an average of 14 fingerprints forms a single cluster). Comparing the cluster classifier to the graph classifier using a paired t-test shows that the graph classifier significantly improves the predictive performance for both the RF and GBM algorithms. For RF, the p-value for the AUC is 0.02, and the p-value for the accuracy is 0.01. For GBM, the p-value for the AUC is 0.04 and 0.002 for accuracy. A comparison of the fingerprint classifier to the cluster classifier shows a statistically significant difference for GBM's accuracy with a p-value of 0.007. The results described above indicate that the graph classifier provides the best performance for both of the algorithms.

Comparing RF to GBM, there is a statistically significant difference for both the fingerprint classifier and the cluster classifier, but there is no significant difference for the graph classifier.

The indoor percentage column represents the accuracy of a constant classifier that classifies all instances as indoors (majority rule); such a classifier has a constant AUC of 0.5. Comparing this constant classifier to the fingerprint classifier shows no significant difference for both RF and GBM accuracy; on the other hand, comparing the AUC of both algorithms demonstrated the superiority of the fingerprint classifier, with a p-value of 3.7E-07 for the RF and 2.88E-8 for the GBM. The variation between the measurements reinforces, once again, the superiority, reliability, and relevance of the AUC measurement in comparison to the accuracy measurement when dealing with data biased toward one class, as in our case where the data is biased toward the indoor samples. Fig. 4 provides a comparison of classification algorithms over all of the datasets, in order to determine which classification algorithm performs best at this task. The figure compares the cluster and transition graph classifiers. The raw fingerprint classifier is not included in this comparison due to its relatively low AUC and accuracy which can be seen in Table 3.

The comparison shows that the ensemble based methods are the most suitable methods for this task. A few other classifiers were tested, including various SVM techniques and neural networks, however they are not presented here due to their poor performance. As can be seen, the naïve Bayes cluster classifier provides very weak performance, and the bagged CART and rotation forest show unusual results in that the graph classifier is significantly better in terms of AUC but not accuracy. The bagged CART and rotation forest classifiers fail to distribute the instances above or under the threshold for classification (0.5 in this case), resulting in relatively low accuracy. Further examination shows that the threshold cannot be reset, as the optimal value is dramatically different for each of the devices. RF and GBM overcome this issue and show improvement for both of the classifiers and measures. We focused on these algorithms to evaluate each device separately at the beginning of this section. GBM usually excels at regression, rather than classification; nevertheless, it shows better results in this experiment.

Fig. 5 shows the detection latency for switches (between indoors and outdoors) for each of the users using a model created from graph features and the GBM classifier. The overall latency for going indoors is 4.3 seconds (based on 82 switches), and 16.6 for going outdoors (based on 93 switches). Additionally, there are three going outdoors switches and two going indoors switches that were not detected (considered as missed), and three switches which had very high latency (>500 seconds) that were also considered as missed. In total, 4.4% of the switches were missed. To complete our comprehensive evaluation, a detailed analysis of the errors was performed. The model's error rate is 6.6%. 7.1% of the errors were caused after a user switched his/her location and before the model detected the switch. 68.3% of the errors are false negatives – fingerprints collected indoors and classified as outdoors, and 31.7% are false positives. More specifically, the majority of errors (64.4%) are false negatives which occurred after correctly detecting the last location change.

## 5.2. Toward an Innovative On-Device Classifier Application

The main drawback of the experiment described above is its offline processing in which the classification process was executed after all of the data was collected and the transition graph was completely constructed. An actual context service should support online classification with a high level of accuracy and confidence and fast turnaround time. In order to assess the feasibility of converting our offline classifier to an online/real-time classifier, we conducted a series of tests on additional devices in new environments, and the results indicate a *warm-up time of a minute* in most cases.

The classifier was trained as previously described, without any tuning to optimize it for online classification. The locations tested were not included in the training set, and the test data was collected by the research team using Samsung S3 (GT-I9300) and Nexus 3 (GT-I9250) devices simultaneously (they were stowed in different pockets for the entire time of the experiment). In order to evaluate the time it took to detect a new location, we constructed a transition graph using the data collected during the first minute and classified the fingerprints. We repeated the process by adding data obtained during each additional minute gradually and evaluated the data collected from the beginning of the test until that minute. This evaluation method allowed us to detect if the classifiers had a "cold start" which is characterized by low performance at the beginning of the test, with an improvement over time. Table 4 describes the scenarios evaluated, each of which is ten minutes long.

The aim of this evaluation is to assess whether the classifier is able to adapt quickly to a new environment and operate accurately there. In order to evaluate the classifier in these *single class* scenarios over time, we only measure accuracy. Figures 6-13 show the accuracy of the cluster and graph based classifiers per minute. For simplicity, we present the results of our evaluation for GBM only, as the RF provides similar results. This experiment also allows us to test if there is a significant difference between the performances of the devices.

Comparing the cluster classifier to the graph classifier shows a significant difference for both the Nexus 3 and Samsung S3 devices, with p-values of 6E-5 and 1.5E-5, respectively. As a result, we focus on the graph classifier in this discussion. Fig. 11 shows the results from the data collected in the underground parking area. The figure reveals a limitation of our method as the classifier mistakenly classifies the vast majority of the instances as "outdoors." The parking lot is an unusual Wi-Fi environment for an indoor area; the scans receive a small number of APs with low signal strength, which explains this discrepancy. After acknowledging this, we omit this scenario later in this discussion.

The first two minutes were classified correctly in all of the scenarios and devices but two: 1) the Nexus 3 in scenario 7 (indoors - shopping mall, Fig. 12), with perfect accuracy in the first minute that drops to a rate of 0.79 in the second minute, and 2) the Samsung S3 in scenario 8 (indoors - classroom building, Fig. 13), which suffers from relatively lower accuracy after the second minute as well. More examples of the warm-up time with the graph classifier can be found in scenario 1 (indoors - office building, Fig. 6), during minutes three to four where the classifier misclassifies a few instances; however, in this case, the classifier classifies the instances correctly in the next minute. This "cold start" behavior is seen more obviously in scenario 7 (indoors - shopping mall, Fig. 12) where the classifier's accuracy increases after two minutes.

Figures 14 and 15 contain box-and-whisker plots for the devices used in scenarios 1-8. It's clear that there are differences between the number of APs detected and the RSSI readings. This can be seen in Fig. 14 for the Samsung device in the sample with the largest number of APs. It corresponds to scenario 8, which suffers from relatively lower accuracy with the Samsung device. The system was trained on a Nexus 5 device; while this might be the reason for the difference between the devices in scenario 8, this difference should be investigated further. Nevertheless, on average the graph classifier accuracy is similar for both of the devices, and a paired t-test for the average accuracy shows no significant difference.

To summarize, the offline model's performance shows a warm-up time of one minute in most cases, and up to three minutes in other cases. Converting our offline classifier to an online version and optimizing it is left for future work.

## 6. Discussion

This section is devoted to discussing the results obtained in the experiments presented, as well as the limitations of this research. Using the large dataset that was collected, our method shows a high average AUC and accuracy, regardless of users' habits (percent of time indoors); for example, the method achieves high performance for users such as user 1 (a student) who spends most of his/her time indoors, as well as user 5 (a business person) who spends about 50% of the time outdoors. We trained our model with a single device and used location based cross-validation to tune it; the model was tested on 12 users. Our users utilized a variety of devices, and yet our method's performance was high regardless of the device used. However, there are some differences which should be further investigated. We demonstrated a simple method based on a linear regression model for selecting the right neighborhood size needed to extract features from the transition graph. Our results confirm that supervised learning based on Wi-Fi fingerprints as instances is suitable for indoor-outdoor detection, and that the construction of the transition graph improves the results significantly. The training and test sets were collected in different locations, which means that the model is able to generalize the knowledge to unfamiliar circumstances.

Random forest and gradient boosting machine algorithms are found to be suitable for classifying Wi-Fi fingerprints in the raw clusters and transition graph representations. Although GBM usually excels with regression problems, it shows the best results in this classification problem among the algorithms tested. It should be mentioned that ensemble methods such as the ones described above are more complex and require more resources to run—a drawback which can affect real-time implementation on a mobile device. The feasibility test for converting the method to an online method suggests a warm-up time of a minute in most cases. However, this test pointed to a specific location where the method has difficulties: the underground parking lot. Additionally, in some cases, such as scenario 8, different device models seem to have some effect on the accuracy. This limitation should be further investigated and is left for future work.

A comprehensive comparison of the available systems for indoor-outdoor detection is presented in Table 5. We report the sensors used by each system, the hardware and site survey requirements of each system, the reported detection performance, and the evaluation method used.

Unfortunately, none of the systems presented in the table report the AUC. Although comparing accuracy between different datasets is not meaningful, for the dataset collected in this experiment we report the accuracy to be 93%, while the AUC is 0.94. In Table 5, the evaluation methods are scored on a scale ranging from low to high based on the following three criteria: (1) the data - real world data, or at least large predefined locations, as opposed to small predefined locations, (2) the number of devices used during evaluation, and (3) the validation technique (10-fold cross-validation vs. device based).

None of the previous research conducted is based on real world data from an experiment involving different users during their daily routine who used various device models, with 828 hours' worth of data, in which the tested method was properly evaluated, as we have done in this research. Anagnostopoulos et al. [8] evaluated their system on a large dataset, but unfortunately the system was evaluated using 10-fold cross-validation which cannot be used to make accurate conclusions as we describe in subsection 4.3.

The indoor-outdoor detection method presented in this research is expected to pave the way for continuous indoor-outdoor detection on mobile devices, which will allow mobile devices to provide better context-based information and services to users. The classification model can be stored on the mobile device to allow anonymous use, or alternatively, the data can be processed on a remote server, a setup which will reduce the load on the mobile devices. The Wi-Fi scan rate can be optimized to reduce the requirements from the mobile device by reducing the scan rate when the device is stationary and increasing it based on the amount of motion, using the accelerometer, for example. The method and insights regarding clustering Wi-Fi fingerprints are expected to improve indoor positioning systems, logical localization systems, and indoor mapping systems that are based on Wi-Fi fingerprints. The Wi-Fi fingerprint transition graph acts as an abstract map of the environment and contains meaningful information that can be used for purposes other than indoor-outdoor detection (e.g., classifying the type of building or neighborhood, etc.). If this graph were constructed continuously, it could help in assessing the distance from a certain location and allow place recognition at the room level.

A large part of the evaluation of the methods developed in this research was performed offline and not in real-time, as would be expected from a context service. Due to the limitations of time and resources, we evaluated the method in only eight defined environments. Including additional scenarios and testing them in real-time might reveal other strengths or weaknesses of our method. Furthermore, the use of clustering algorithms other than the light variation of DBSCAN (DBSCANLight) might generate different results.

## 7. Conclusions

In this research, we presented the design, implementation, and evaluation of an indoor-outdoor environment detection method based solely on Wi-Fi fingerprints. Our main idea was to create a transition graph from clusters based on a Wi-Fi fingerprint. This research uses a well-defined rank correlation coefficient, the Spearman correlation, and formulates a distance measure from it for Wi-Fi fingerprint clustering. Our motivation to use Wi-Fi fingerprints is based on the fact that many IPSs are based on Wi-Fi fingerprints, making it highly likely that the Wi-Fi receiver will be on most of the time. The proposed method was implemented and evaluated in an experiment involving 12 participants. A total of 828 hours were logged in this experiment in which our method achieved an AUC score of 0.94 and accuracy of 0.93. A few modifications and improvements are planned for future work. First and foremost is the ability to classify an "instance" in real-time on a mobile device. In addition, the method presented in the current paper can be easily converted for use with GSM fingerprints, and a combination of GSM and Wi-Fi fingerprints might yield better results. Future research should thoroughly investigate the limitations of the method, including the underground parking lot scenario, areas not covered by Wi-Fi Aps, and power consumption optimization. We believe that other context detection tasks can be accomplished using the cluster transition graph, such as classifying the building type (residential, office, etc.), and potentially, the outdoor environment (urban/rural area), which can help create maps by processing Wi-Fi fingerprints and the locations collected by a large number of users.

**Tables**

| User | Device Model | Hours | Total FP | Unlabeled FP | Indoors FP | Outdoors FP | Indoors % |
|---|---|---|---|---|---|---|---|
| 0 | Nexus 5 | 22 | 10230 | 168 | 8128 | 3694 | 69 |
| 1 | Samsung Galaxy S5 | 144 | 55668 | 1765 | 54919 | 1284 | 98 |
| 2 | LG Nexus 5 | 173 | 63736 | 2522 | 60395 | 3852 | 94 |
| 3 | LG G3 | 108 | 44648 | 12040 | 26417 | 7684 | 77 |
| 4 | Samsung Galaxy S2 | 106 | 50452 | 2202 | 43260 | 4996 | 90 |
| 5 | LG Nexus 4 | 31 | 18182 | 0 | 8984 | 9342 | 49 |
| 6 | Samsung Galaxy S3 | 24 | 20087 | 220 | 17646 | 2222 | 89 |
| 7 | LG Nexus 5 | 46 | 39271 | 3027 | 32641 | 3660 | 90 |
| 8 | LG Nexus 4 | 22 | 16799 | 0 | 14160 | 2639 | 84 |
| 9 | LG G3 | 58 | 15896 | 1759 | 13796 | 936 | 94 |
| 10 | Samsung Galaxy S3 | 30 | 25663 | 0 | 24754 | 909 | 96 |
| 11 | LG G3 | 35 | 8239 | 41 | 6714 | 1784 | 79 |
| 12 | Samsung Galaxy S6 | 51 | 24009 | 150 | 16522 | 7337 | 69 |

| | Average | 65 | 30222 | 1838 | 25257 | 3872 | 83 |

**Table 1 - Summary of the data collected from each participant**

| Name | Minimum neighborhood size | Maximum neighborhood size |
|---|---|---|
| Number of neighbors | 2 | 6 |
| Average power in the neighborhood (RSSI) | 0 | 4 |
| Average number of APs per scan in the neighborhood | 0 | 4 |
| Average number of fingerprints in the neighborhood | 0 | 4 |

**Table 2 - Neighborhood size used for each node feature**

| | | | | | | Random Forest | | | | | | Gradient Boosting Machine | | | | | |
|---|---|---|---|---|---|---|---|---|---|---|---|---|---|---|---|---|---|
| | | | | | | FP | | Clusters | | Graph | | FP | | Clusters | | Graph | |
| User | Device Model | Hours | Indoors % | Total FP | Clusters | AUC | ACC | AUC | ACC | AUC | ACC | AUC | ACC | AUC | ACC | AUC | ACC |
| 1 | Samsung Galaxy S5 | 144 | 98 | 55668 | 1330 | 0.84 | 0.97 | 0.86 | 0.99 | 0.96 | 0.99 | 0.94 | 0.99 | 0.96 | 0.99 | 0.97 | 0.99 |
| 2 | LG Nexus 5 | 173 | 94 | 63736 | 4669 | 0.76 | 0.94 | 0.84 | 0.95 | 0.87 | 0.97 | 0.81 | 0.94 | 0.85 | 0.95 | 0.86 | 0.97 |
| 3 | LG G3 | 108 | 77 | 44648 | 4619 | 0.72 | 0.82 | 0.66 | 0.80 | 0.97 | 0.81 | 0.86 | 0.82 | 0.96 | 0.80 | 0.97 | 0.81 |
| 4 | Samsung Galaxy S2 | 106 | 90 | 50452 | 2655 | 0.62 | 0.79 | 0.69 | 0.78 | 0.96 | 0.95 | 0.68 | 0.88 | 0.78 | 0.91 | 0.89 | 0.91 |
| 5 | LG Nexus 4 | 31 | 49 | 18182 | 1566 | 0.88 | 0.80 | 0.95 | 0.92 | 0.97 | 0.95 | 0.90 | 0.86 | 0.98 | 0.92 | 0.99 | 0.97 |
| 6 | Samsung Galaxy S3 | 24 | 89 | 20087 | 2230 | 0.89 | 0.94 | 0.94 | 0.95 | 0.98 | 0.97 | 0.94 | 0.94 | 0.98 | 0.95 | 0.99 | 0.97 |
| 7 | LG Nexus 5 | 46 | 90 | 39271 | 2491 | 0.90 | 0.92 | 0.92 | 0.96 | 0.98 | 0.99 | 0.96 | 0.94 | 0.98 | 0.96 | 0.98 | 0.97 |
| 8 | LG Nexus 4 | 22 | 84 | 16799 | 1784 | 0.80 | 0.84 | 0.77 | 0.77 | 0.83 | 0.91 | 0.85 | 0.85 | 0.87 | 0.87 | 0.89 | 0.91 |
| 9 | LG G3 | 58 | 94 | 15896 | 1609 | 0.85 | 0.87 | 0.87 | 0.94 | 0.90 | 0.97 | 0.88 | 0.93 | 0.94 | 0.94 | 0.95 | 0.97 |
| 10 | Samsung Galaxy S3 | 30 | 96 | 25663 | 765 | 0.96 | 0.96 | 0.95 | 0.99 | 0.95 | 0.98 | 0.98 | 0.98 | 0.95 | 0.99 | 0.95 | 0.98 |
| 11 | LG G3 | 35 | 79 | 8239 | 896 | 0.73 | 0.80 | 0.73 | 0.79 | 0.78 | 0.84 | 0.86 | 0.79 | 0.76 | 0.82 | 0.77 | 0.87 |
| 12 | Samsung Galaxy S6 | 51 | 69 | 24009 | 2572 | 0.71 | 0.67 | 0.79 | 0.64 | 0.79 | 0.71 | 0.72 | 0.66 | 0.76 | 0.70 | 0.86 | 0.73 |
| | **Average** | **65** | **83** | **31887** | **2265** | **0.81** | **0.86** | **0.83** | **0.87** | **0.91** | **0.92** | **0.87** | **0.88** | **0.90** | **0.90** | **0.92** | **0.92** |

**Table 3 - Results based on device used**

| Scenario ID | Scenario Description | Indoors/Outdoors |
|---|---|---|
| 1 | Walking/sitting (alternately) in an office building | Indoors |
| 2 | Walking/standing (alternately) at a business park | Outdoors |
| 3 | Walking/standing (alternately) in a low density urban area | Outdoors |
| 4 | Riding in a car through an urban area | Outdoors |
| 5 | Walking at an open-air food market | Outdoors |
| 6 | Walking/standing (alternately) in an underground parking lot | Indoors |
| 7 | Walking/standing (alternately) in a shopping mall | Indoors |
| 8 | Walking/standing (alternately) in a classroom building | Indoors |

**Table 4 - Description of the tested scenarios**

| System | Sensors used | Can operate without site survey? | Can operate without additional hardware? | Target attribute cardinality and values | Reported detection performance | Strength of evaluation |
|---|---|---|---|---|---|---|
| Proposed system | Wi-Fi | Yes | Yes | 2, indoor & outdoor | AUC of 94%, accuracy of 93% | High. Twelve participants during daily routine (828 hours in total). Seven device models. Trained model using a single device. |
| BlueDetect [3] | Bluetooth | Yes | No | 3, indoor, semi-outdoor, outdoor | Accuracy of 96.2% | Low. Two device models. Single predefined location. |
| Wang et al. [4] | Cell tower RSS | Yes | Yes | 4, open outdoors, semi-outdoors, light indoors, deep indoors | Accuracy of 100% | Low. One device model. Four predefined environments, 10 minutes each. |
| Cho et al. [6] | GPS | Yes | Yes | 2, indoor & outdoor | Presents FPR & FNR graphs | Low. One device model. Single predefined location. |
| IODetector [7] | Accelerator, proximity, light, time, cell tower RSS, magnetic field | Yes | Yes | 3, indoor, semi-outdoor, outdoor | Accuracy of 82%, precision & recall above 88% for each of the three classes | High. Three device models. 19 traces including 84 different sites over one-month period. |
| Anagnostopoulos et al. [8] | Subset of: activity recognition (Google API), barometric pressure, light, proximity, cloud coverage, time, cell tower RSS, accelerometer, magnetometer, microphone, GPS, screen on/off, Wi-Fi | No | Yes | 2, indoor & outdoor | Up to 99% accuracy | Medium. Eleven participants with personal phones. Total of 388 hours of data. 10-fold cross-validation. |
| Radu et al. [9] | Subsets of: light, time, proximity, battery temperature, microphone, cell tower RSS, magnetometer | Yes | Yes | 2, indoor & outdoor | Accuracy greater than 90% | Low. Two device models. Three predefined environments. |
| WifiBoost [10] | Wi-Fi | No | Yes | 2, indoor & outdoor | Error rate of 2% | Low. Two device models. Single predefined location. |
| Edelev et al. [11] | GPS, time, microphone, temperature, humidity, network connectivity | Yes | Yes | 2, indoor & outdoor | Accuracy of 92.7% | Low. Three devices. Data collected in and between various buildings on a campus. |
| He, Tan and Gary Chan [12] | Wi-Fi | No | Yes | 2, indoor & outdoor | Accuracy of 95.69%, TPR of 99.73%, TNR of 95.6% | High. More than four device models. Four predefined large locations. Over 45,000 fingerprints. |

| Anagnostopoulos and Deriaz [13] | Any positioning technology that supports error estimation | Yes | Yes | 2, indoor & outdoor | Delay of switching of 3-9 seconds | Low. One device model. Single predefined location. |

**Table 5 - Comparison of indoor-outdoor detection systems**

# Figures

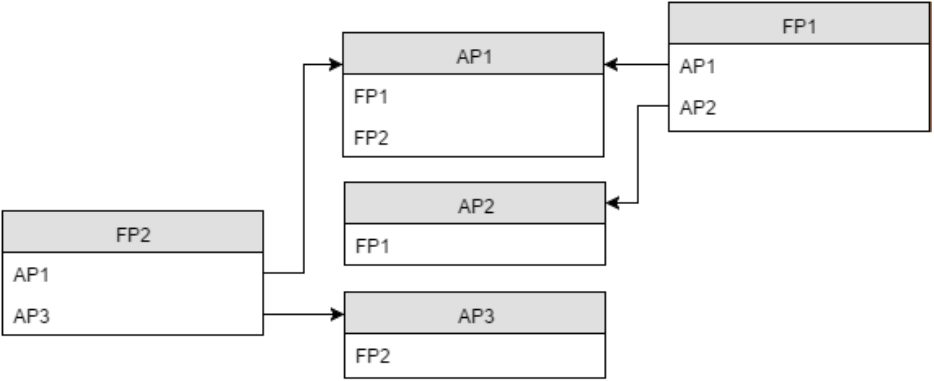

**Fig. 1. Fingerprint to AP and AP to fingerprint mapping**

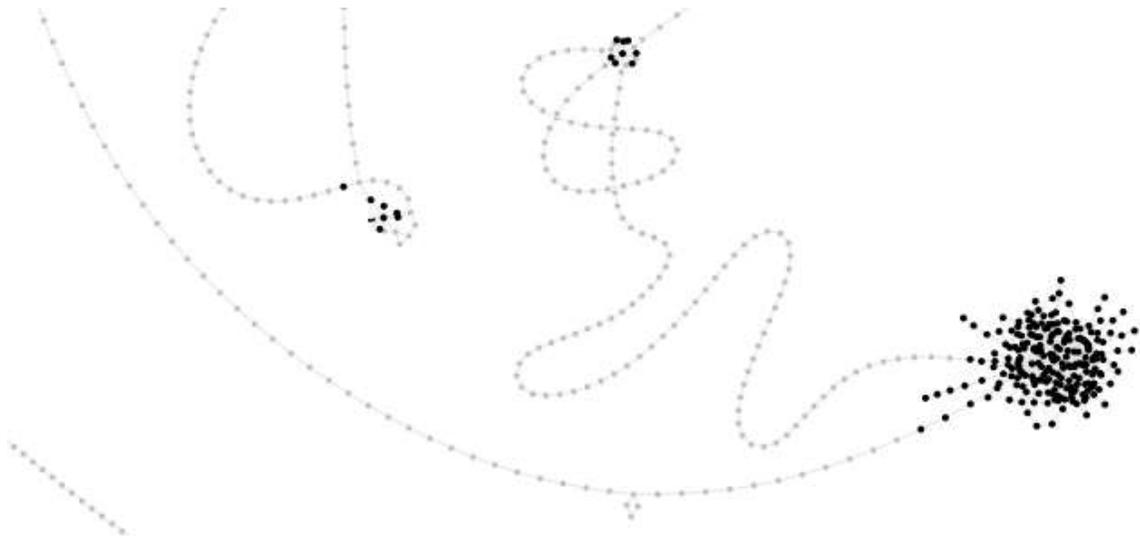

**Fig. 2. Transition graph (darker nodes were collected indoors, and lighter nodes were collected outdoors)**

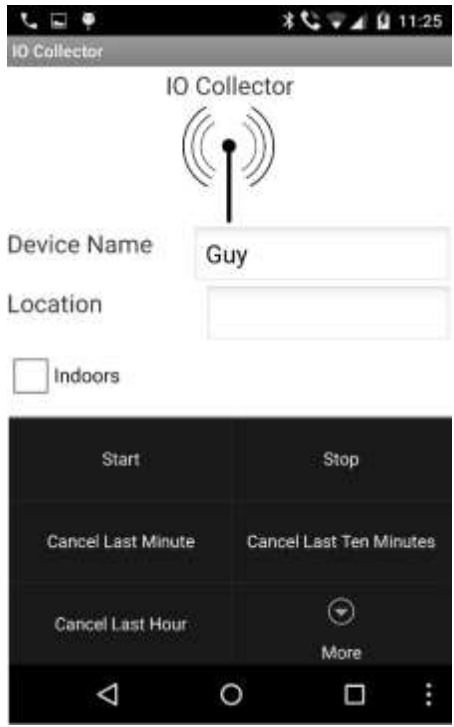

**Fig. 3. The Android application for data collection used in the experiment**

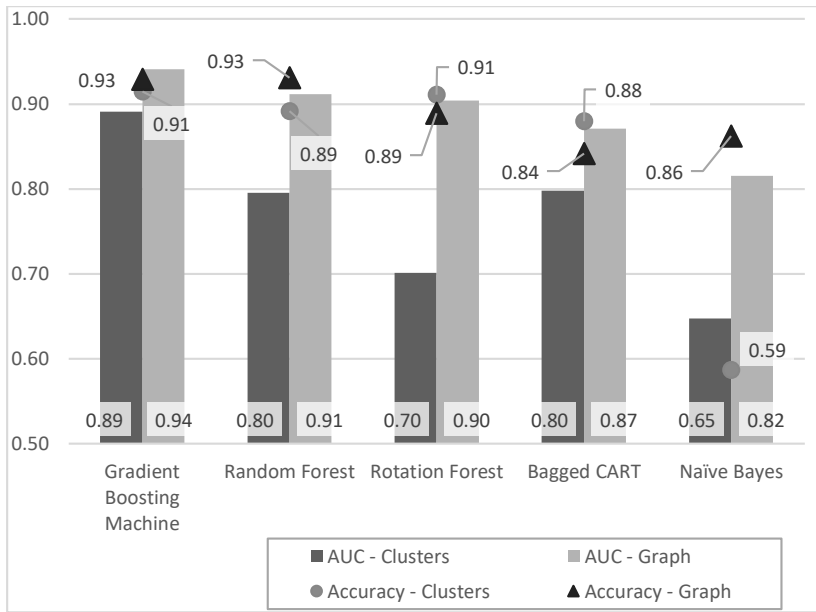

**Fig. 4. Comparison between various classifiers for all of the devices**

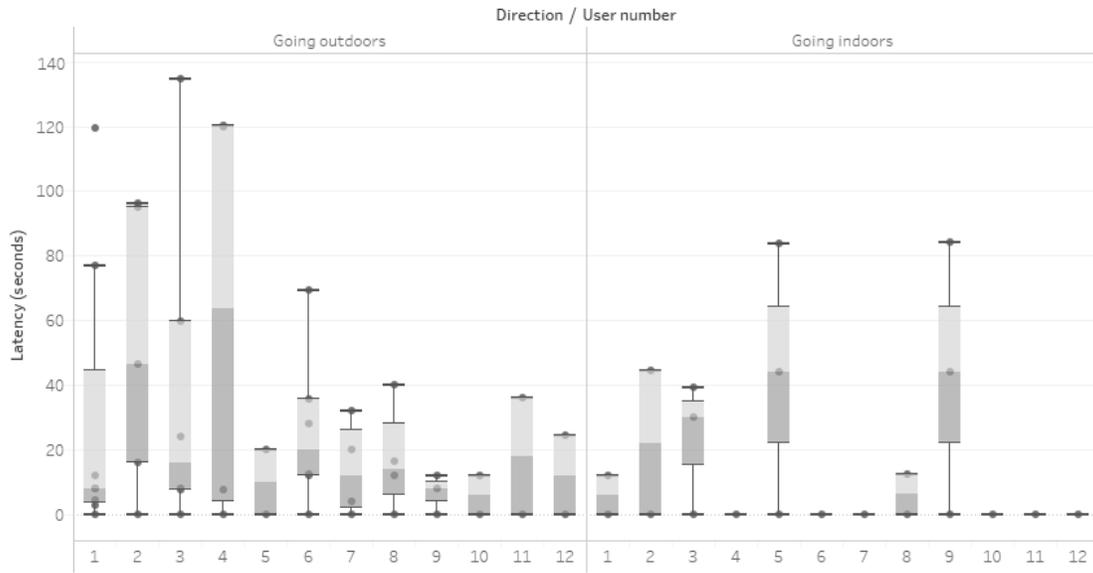

**Fig. 5. The detection latency for going indoors and outdoors (per user)**

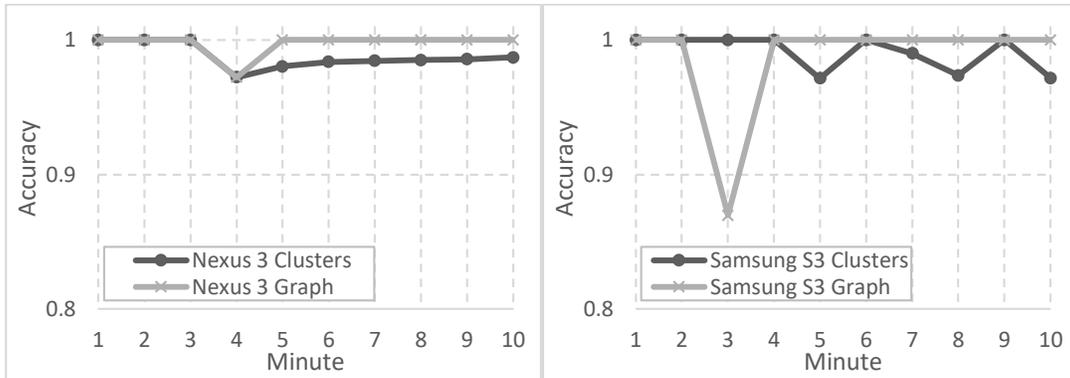

**Fig. 6. Per minute accuracy for scenario 1 (indoors - office building)**

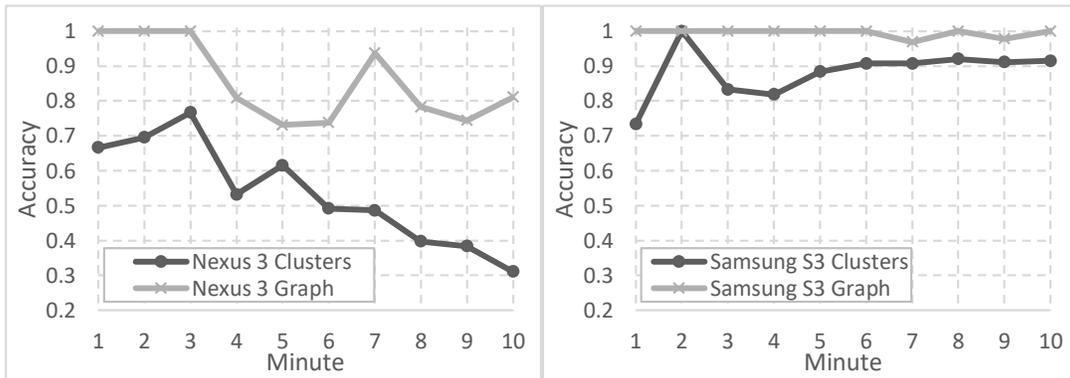

**Fig. 7. Per minute accuracy for scenario 2 (outdoors - business park)**

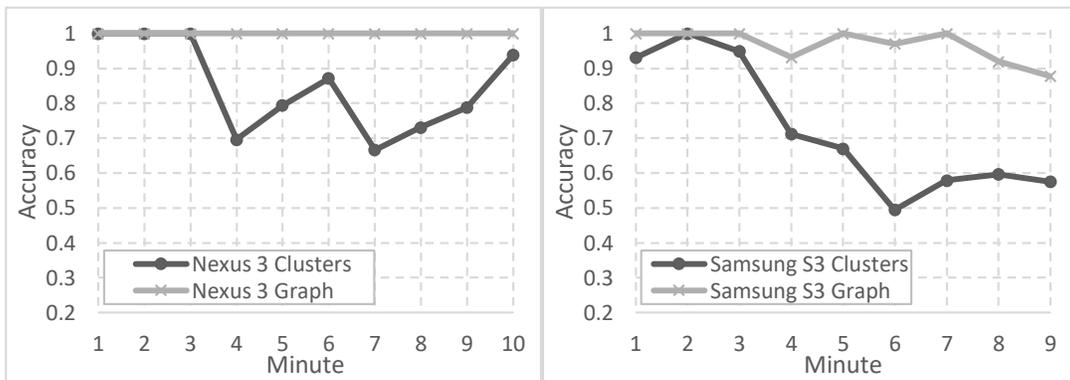

**Fig. 8. Per minute accuracy for scenario 3 (outdoors - low density urban area)**

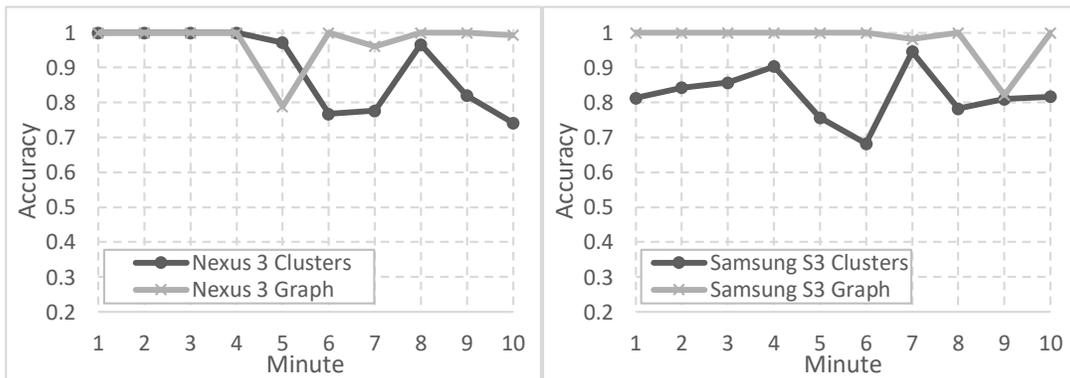

**Fig. 9. Per minute accuracy for scenario 4 (outdoors - car in urban area)**

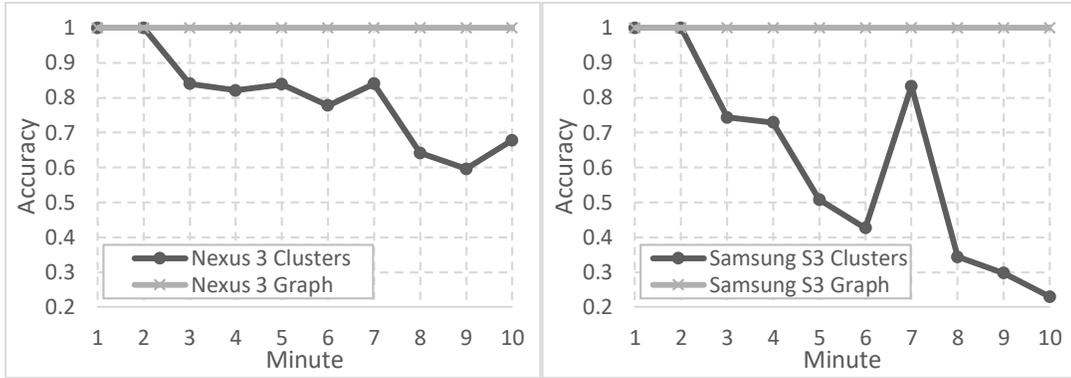
Fig. 10. Per minute accuracy for scenario 5 (outdoors - open-air food market)

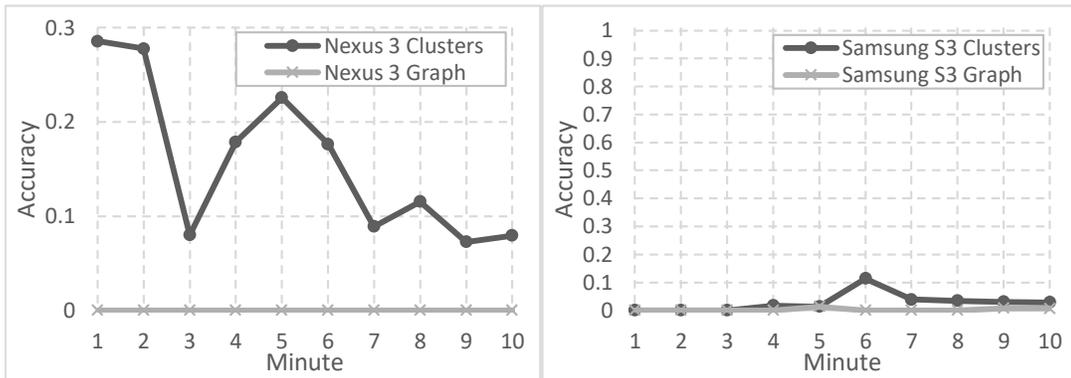
Fig. 11. Per minute accuracy for scenario 6 (indoors - underground parking lot)

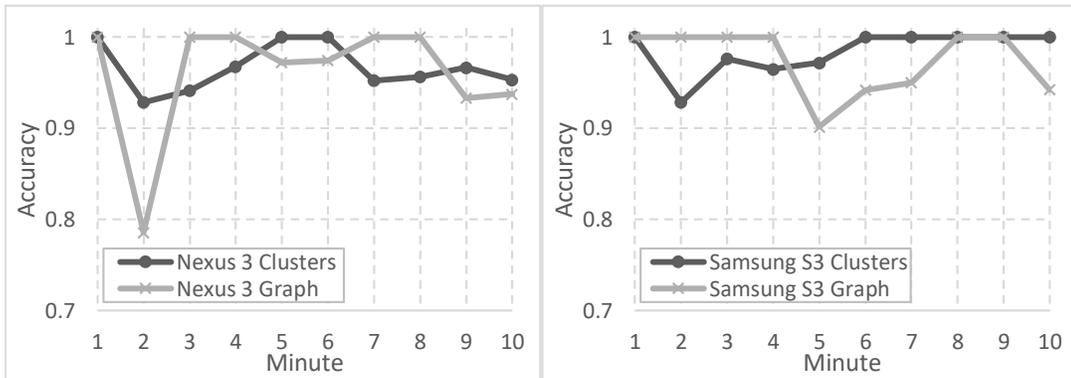
Fig. 12. Per minute accuracy for scenario 7 (indoors - shopping mall)

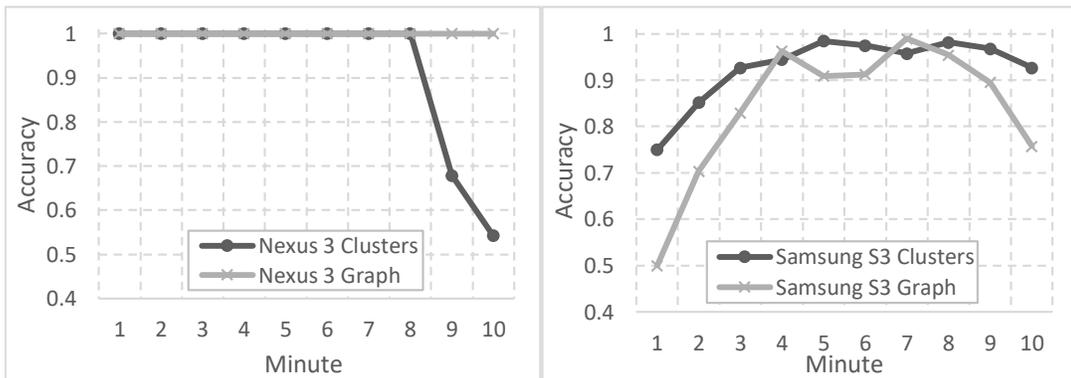
Fig. 13. Per minute accuracy for scenario 8 (indoors - classroom building)

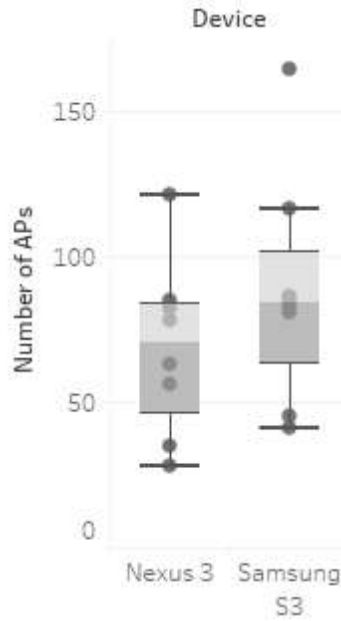

**Fig. 14. A comparison of the number of APs detected in scenarios 1-8**

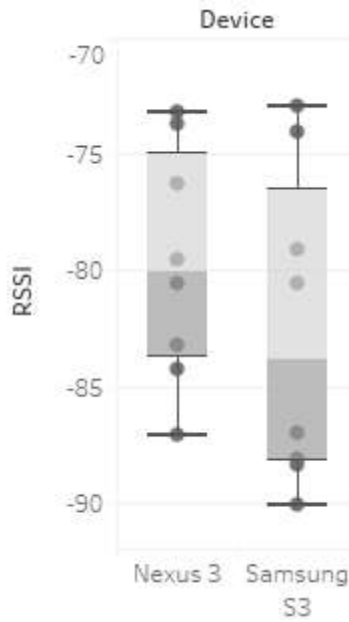

**Fig. 15. A comparison of the RSSI readings in scenarios 1-8**